\documentclass[a4paper]{article}

\usepackage{itgspeech2025}    
\usepackage{multirow}
\usepackage{times}            
\usepackage[english]{babel}   
\usepackage[ansinew]{inputenc}
\usepackage[T1]{fontenc}      
\usepackage[sort&compress,numbers]{natbib}	
\usepackage{amsmath,amssymb}
\usepackage{graphicx}
\usepackage[colorlinks=false,pdfborder={0 0 0}]{hyperref}
\usepackage{booktabs}
\usepackage{acronym}
\usepackage{tikz, tikz-qtree, units}
\usepackage{balance}
\usepackage{mathabx}

\title{Low-Complexity Neural Wind Noise Reduction for Audio Recordings}

\author{Hesam Eftekhari$^1$, Srikanth Raj Chetupalli$^2$, Shrishti Saha Shetu$^2$, Emanu\"{e}l A.~P.~Habets$^2$, and Oliver Thiergart$^2$}
\address{\textit{$^1$ Fraunhofer IIS, Am Wolfsmantel 33, 91058 Erlangen, Germany}\\
\textit{$^2$ International Audio Laboratories Erlangen\thanks{A joint institution of Fraunhofer IIS and Friedrich-Alexander-Universit{\"a}t Erlangen-N{\"u}rnberg (FAU), Germany.
}, Am Wolfsmantel 33, 91058 Erlangen, Germany
} \\
Emails: \small \textit{\{hesam.eftekhari, 
srikanth.chetupalli, shrishti.saha.shetu, emanuel.habets, oliver.thiergart\}@iis.fraunhofer.de}}

\acrodef{WNR}{wind noise reduction}
\acrodef{DNN}{deep neural network}
\acrodef{LSTM}{long short-term memory}
\acrodef{PSD}{power spectral density}
\acrodef{WNE}{wind noise estimation}
\acrodef{STFT}{short-time Fourier transform}
\acrodef{SNR}{signal-to-noise ratio}
\acrodef{SE}{speech enhancement}

\begin{document}

\maketitle
\sloppy
\begin{abstract}
Wind noise significantly degrades the quality of outdoor audio recordings, yet remains difficult to suppress in real-time on resource-constrained devices. In this work, we propose a low-complexity single-channel deep neural network that leverages the spectral characteristics of wind noise. Experimental results show that our method achieves performance comparable to the state-of-the-art low-complexity ULCNet model. The proposed model, with only $249$K parameters and roughly $73$~MHz of computational power, is suitable for embedded and mobile audio applications.
\end{abstract}

\footnote{The authors thank the Erlangen Regional Computing Center~(RRZE) for providing compute resources and technical support.}
\vspace{-0.5em}
\section{Introduction}
\label{sec:intro}

A microphone captures a range of acoustic sources present in the scene. In an outdoor setting, it may also pick up noise induced by the wind flowing past it. Devices such as mobile phones, action cameras, wearables, and hearables frequently encounter wind noise. The turbulent nature of wind flow results in a high-amplitude signal at low frequencies, leading to undesirable distortions in the recorded signal. Hence, effective \ac{WNR} has become essential to modern consumer audio devices. The increasing adoption of wearables and hearables further amplifies the demand for low-complexity \ac{WNR} solutions. Although the \ac{WNR} task resembles \ac{SE}, it differs in its objective: the target signal consists of all sound sources, excluding the wind noise component.

Several single-channel approaches \cite{5346518,6333897,7362506, 6854970, 7895558, 8456465, 9042991, 10509629} and multi-channel methods \cite{franz2010multi, 7776196, 9103994} utilizing traditional signal processing and \ac{DNN}-based techniques have been proposed. Earlier approaches used techniques such as adaptive post-filtering in the time domain \cite{5346518} and methods derived from artificial bandwidth extension \cite{6333897}. Other approaches \cite{7362506, 6854970} concentrated on the estimation of the noise \ac{PSD}, which is used to obtain spectral gains for \ac{WNR}. Wind noise exhibits distinct statistical properties in comparison to other acoustic signals, which are leveraged for the \ac{PSD} estimation. In this context, multi-microphone approaches \cite{franz2010multi, 7776196, 9103994}, which exploit the spatial characteristics of the received wind noise, have proven to be successful. 

In the deep learning context, a bi-directional \ac{LSTM}-based approach was first proposed in \cite{7895558}. In \cite{8456465}, the authors investigated the estimation of separate masks for the wind and desired signal components, followed by soft masking based on the auditory masking properties. More recent studies \cite{9042991, 10509629} employed a U-Net-based architecture \cite{UNet} composed of 2D convolutional layers. However, these \ac{DNN}-based models are not designed for frame-by-frame processing and do not meet the memory and computational complexity constraints of embedded hardware. On the other hand, the \ac{SE} literature includes several methods that satisfy such design requirements (e.g., \cite{ULCNet_24}); however, these approaches are typically optimized for speech-only scenarios and often suppress music and other non-speech acoustic sources.

In this work, we focus on low-complexity single-channel \ac{WNR} techniques for real-time applications on embedded platforms using \acp{DNN}. We aim to directly estimate the desired signal by computing a mask, referred to here as the \textit{wind rejection} approach. Since the desired signal in \ac{WNR} includes all acoustic sources except wind noise, we also investigate a \textit{wind extraction} approach, where the \ac{DNN} estimates a mask for the wind noise itself, which is then removed from the microphone signal to obtain the desired signal.
Wind noise is generally concentrated in low-frequency regions with negligible presence in frequencies above $4$ kHz. Motivated by this observation, we propose a dual encoder architecture that emphasizes low-frequency bands while focusing less on higher frequency bands. This design is integrated into the low-complexity ULCNet architecture originally proposed for \ac{SE} task \cite{ULCNet_24}.
Experimental results using speech and music sources validate the effectiveness of the proposed approach for \ac{WNR} across varied acoustic scenes.
\vspace{-0.5em}
\section{Problem Formulation}
\label{sec:problem_formulation}
Consider a recording scenario where a single microphone captures an acoustic scene comprising arbitrary sound sources and wind noise. The signal model is expressed as,
{\small
\begin{equation} 
    \begin{split} 
        \mathbf{x}[n]
                = \underbrace{ \mathbf{s}[n] + \sum_{j=1}^{I_b} \mathbf{b}_j[n]}_{\mathbf{d}[n]} + \mathbf{w}[n],
    \end{split}
\label{eq:extended_signal_model}
\end{equation}
}
where \( \mathbf{x}[n] \) denotes the recorded microphone signal at discrete time index \( n \), composed of the desired signal \( \mathbf{d}[n] \) and additive wind noise \( \mathbf{w}[n] \). The desired signal \( \mathbf{d}[n] \) may consist of speech components \( \mathbf{s}[n] \) and \( I_b \) non-speech components \( b_j[n] \), representing music, ambient noise, or other acoustic sources. 

In conventional DNN-based \ac{SE} approaches \cite{ULCNet_24,shetu2024comparative}, a DNN \( \mathcal{F}_\theta \) is trained to estimate the clean speech signal \( \widehat{\mathbf{s}}[n] \) from the observed noisy signal \( \mathbf{x}[n] \) using mapping, masking, or deep filtering methods \cite{zheng2023sixty}. 
This formulation does not account for the presence of other non-speech components \(\sum_{j=1}^{I_b} \mathbf{b}_j[n]\). 
In this work, we define wind noise as the undesired signal, while treating all other signal components comprising speech or non-speech sound sources as desired, i.e., given the noisy signal \({\bf x}[n]\), the goal in \ac{WNR} is to estimate the desired signal \({\bf d}[n]\). 

We formulate the wind noise reduction (WNR) problem using two distinct paradigms: \textit{wind rejection}  and \textit{wind extraction}. In \textit{wind rejection}, a DNN denoted as \(\mathcal{F}_{\textrm{rej}}\) directly estimates the desired signal,
{\small
\begin{equation}\label{eqn:wind_rej}
\widehat{{\bf d}}[n] = \mathcal{F}_{\textrm{rej}}({\bf x}[n]). 
\end{equation}
}
However, this approach, especially when combined with masking-based desired signal reconstruction method, often results in signal distortions, particularly in low \ac{SNR} scenarios where \(\mathbb{E}\left[\|\mathbf{d}[n]\|^2\right] \ll \mathbb{E}\left[\|\mathbf{w}[n]\|^2\right]\)
 \cite{shetu2024comparative,hao2020masking}. Therefore, we alternatively investigate a two-stage approach, referred to as \textit{wind extraction}. In the first stage, a DNN, denoted by \( \mathcal{F}_{\textrm{ext}} \), estimates the wind noise component
 {\small
\begin{equation}\label{eqn:windExtr}
\widehat{{\bf w}}[n] = \mathcal{F}_{\textrm{ext}}({\bf x}[n]).
\end{equation}
}
Subsequently, the desired signal is recovered using traditional signal processing methods such as post-filtering \cite{5346518}, or direct subtraction \cite{4146534} as
{\small
\begin{equation}\label{eqn:windExtr2}
\widehat{{\bf d}}[n] = {\bf x}[n] - \widehat{{\bf w}}[n].
\end{equation}
}
\section{Proposed Method}
\label{sec:proposed_solution}
\begin{figure}[t!]
    \centering
    \begin{tikzpicture}[scale=0.9, transform shape]
        \node[] (x) at (-1.5,-0.5) {${\bf x}$};
        \node[draw=red, text width=0.8in, minimum height=2.4em, align=center, rounded corners] (ip) at (-1.5,-2) {Input preprocessing};
        \node[draw=red, text width=0.8in, minimum height=2.5em, align=center, rounded corners] (cwr) at (-1.5,-4) {Channel-wise feature reorientation };    
        \node[draw=red, text width=0.5in, minimum height=2.5em, align=center, rounded corners] (lfe) at (-2.3,-6) {LF encoder};
        \node[draw=red, text width=0.5in, minimum height=2.5em, align=center, rounded corners] (hfe) at (-0.7,-6) {HF encoder};

        \node[draw=red, text width=0.7in, minimum height=2.5em, align=center, rounded corners] (im) at (1.2,-2) {Intermediate mask};
        \node[draw=red, text width=0.7in, minimum height=2.5em, align=center, rounded corners] (fc) at (1.2,-3.35) {Fully Connected};
        \node[draw=red, text width=0.5in, minimum height=2.5em, align=center, rounded corners] (con) at (1.2,-4.65) {Concat};
        \node[draw=red, text width=0.5in, minimum height=2.5em, align=center, rounded corners] (gru) at (1.2,-6) {GRU};     

        \node[draw=blue, align=center,circle, inner sep=1pt] (ss) at (3.8,-0.5) {$-$};    
        \node[draw=blue, text width=0.6in, minimum height=2.5em, align=center, rounded corners] (iplc) at (3.8,-2) {inverse power-law};
        \node[draw=blue, text width=0.6in, minimum height=2.5em, align=center, rounded corners] (mask) at (3.8,-3.35) {Masking};
        \node[draw=blue, text width=0.6in, minimum height=2.5em, align=center, rounded corners] (pc) at (3.8,-4.65) {Pointwise Conv};
        \node[draw=blue, text width=0.6in, minimum height=2.5em, align=center, rounded corners] (cnn) at (3.8,-6) {CNN};

        \draw[->] (x) -- (ip);
        \draw[->] (ip) --node[left]{$\widetilde{\mathbf{X}}_{m}$} (cwr);
        \draw[->] ([xshift=-0.8cm]cwr.south) --node[left]{\(\widetilde{\mathbf{X}}_{c\text{-low}}\)} (lfe);
        \draw[->] ([xshift=0.8cm]cwr.south) --node[left]{\(\widetilde{\mathbf{X}}_{c\text{-high}}\)} (hfe);
        
        \draw[->] (lfe.south) -- ([yshift=-0.25cm]lfe.south) --node[below, xshift=1.9cm]{$\mathbf{C}_{l}$}([yshift=-0.25cm, xshift=3.5cm]lfe.south) -- (gru.south);
        \draw[->] (hfe.east) -- ([xshift=0.15cm]hfe.east) --node[left]{$\mathbf{C}_h$} ([yshift=1.35cm, xshift=0.15cm]hfe.east) -- (con.west);
        \draw[->] (gru) --node[left]{$\mathbf{H}_l$} (con);
        \draw[->] (con) -- (fc);
        \draw[->] (fc) --node[left]{$\widebar{\mathbf{M}}_{m}$} (im);
        \draw[->] (ip) --node[above]{$\widetilde{\mathbf{X}}_{p}$} (im);
        \draw[dashed, red] (-3.5,-1.2) rectangle (2.4,-7.2);
        \draw[dashed, blue] (2.6,-1.2) rectangle (5.0,-7.2);
        \node[] at (-0.4,-7.0) {First stage};
        \node[] at (3.8,-7.0) {Second stage};  
        
        \draw[->] (im.east) -- ([xshift=0.1cm]im.east) |- (cnn.west);
        \draw[->] ([xshift=0.8cm]ip.north) -- ([yshift=0.15cm, xshift=0.8cm]ip.north) --node[above]{$\left[\widetilde{\mathbf{X}}_{r}, \widetilde{\mathbf{X}}_{i}\right]$} ([yshift=0.15cm, xshift=4.0cm]ip.north) |- (mask.west);
        
        \draw[->] (cnn) -- (pc);        
        \draw[->] (pc) -- (mask);
        \draw[->] (mask) -- (iplc);
        \draw[->] (iplc) -- (ss);
        \draw[->] (x) -- (ss);
        \draw[->] (ss.east) --node[right, pos=1.1]{$\hat{\bf d}$} ([xshift=0.5cm]ss.east);
    \end{tikzpicture}
    \vspace{-1em}
\caption{The proposed \textit{WindNetLite} model.}
    \label{fig:bd}
        \vspace{-1.5em}
\end{figure}
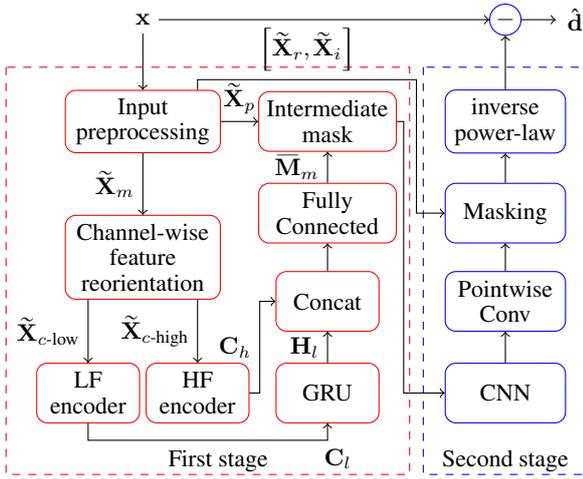
In our work, we consider a real-time audio application scenario that requires causal and low-complexity DNN processing. To this end, we adopt the ULCNet-based model architecture~\cite{ULCNet_24}, originally proposed for noise reduction and shown to be effective in related tasks, such as acoustic echo cancellation and dereverberation~\cite{shetu2024hybrid,shetu2024align,rao2025low}. ULCNet uses a two-stage process of first estimating a real-valued magnitude mask and then refining it to a phase-enhanced complex mask. We modify this architecture by adding two parallel encoder blocks focusing on low and high frequency regions, respectively, to extract separate intermediate features by exploiting wind noise characteristics. The resulting architecture, referred to as \(\textrm{\emph{WindNetLite}}\), is shown in Fig.~\ref{fig:bd}. In the following, we describe the \(\textrm{\emph{WindNetLite}}\) architecture in detail. 
\subsection{DNN Processing}

\begin{figure}[t!]
\centering
\hspace{-0.5cm}
\input{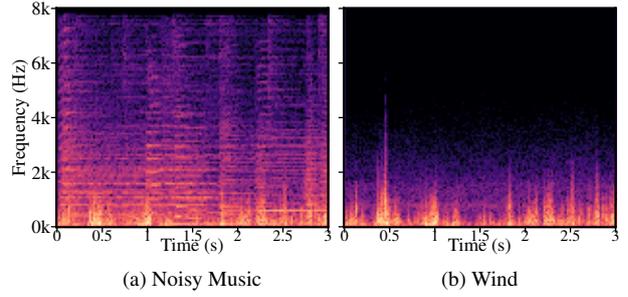}
\vspace{-1.5em}
\caption{ (a) A Jazz music signal, corrupted by strong (b) wind noise, limited to the $2-3$ kHz frequency range.}
\label{fig:windsignal}
\vspace{-1.5em}
\end{figure}
In our architecture, the input preprocessing block first transforms the time-domain signal \(\mathbf{x}[n]\) to \ac{STFT} domain $\mathbf{X}$ and then into a power-law compressed representation {\small$\widetilde{\mathbf{X}}$}, and extracts magnitude and phase features, as described in \cite{ULCNet_24},
{\small
\begin{equation}
    \widetilde{\mathbf{X}} = \operatorname{sign}(\Re(\mathbf{X}))  |\Re(\mathbf{X})|^{\alpha} + j \cdot \operatorname{sign}(\Im(\mathbf{X}))  |\Im(\mathbf{X})|^{\alpha},
    \label{eq:powerlaw_real_imag}
\end{equation}
}
{\small
\begin{equation}
    \widetilde{\mathbf{X}}_{m} = \sqrt{\widetilde{\mathbf{X}}_{r}^2 + \widetilde{\mathbf{X}}_{i}^2}, \quad
    \widetilde{\mathbf{X}}_{p} = \arctan\left( \frac{\widetilde{\mathbf{X}}_{i}}{\widetilde{\mathbf{X}}_{r}} \right).
    \label{eq:mag_phase}
\end{equation}
}
Here, \( j \) represents  \(\sqrt{-1} \), \(\alpha \in [0,1]\) denotes the power-law compression factor, and the \(\operatorname{sign}\) operator ensures the original sign of the real and imaginary parts is preserved in \(\widetilde{\mathbf{X}}_{r}\) and \(\widetilde{\mathbf{X}}_{i}\). The magnitude feature \(\widetilde{\mathbf{X}}_{m} \in \mathbb{R}^{B \times T \times F}\) is further reoriented along the channel dimension to reduce the frequency dimension, which aids in reducing the computational cost of convolutional operations~\cite{ULCNet_24}. The reoriented channel-wise representation is \(\widetilde{\mathbf{X}}_{c} \in \mathbb{R}^{B \times T \times L \times N}\), and considering only the frequency axis, it is defined as:
{\small
\begin{equation}
    \widetilde{\mathbf{X}}_c = \text{stack}\left( \left\{ \widetilde{\mathbf{X}}_m[\ldots,i \cdot \widetilde{L} : i \cdot \widetilde{L} + L] \right\}_{i=0}^{N-1} \right),
\end{equation}
}
where \(L\) is the sub-band length, \(N\) is the total number of sub-bands, \(r\) is the overlap ratio,  and \(\widetilde{L}\triangleq L (1 - r)\). 

Wind, as illustrated in Fig.~\ref{fig:windsignal}(b), typically exhibits high energy in the low-frequency range and negligible energy at frequencies higher than $4$~kHz. Motivated by this, we propose a dual-encoder architecture with separate encoders for low- and high-frequency regions. The magnitude features \(\widetilde{\mathbf{X}}_c\) are first split into low- and high-frequency components, where \(\widetilde{\mathbf{X}}_{c\text{-low}}\) contains the first 5 sub-bands, and \(\widetilde{\mathbf{X}}_{c\text{-high}}\) contains the remaining sub-bands. These components are processed in parallel by two encoder blocks, termed the \textrm{LF encoder} and \textrm{HF encoder}, each consisting of convolutional layers followed by batch normalization and ReLU activation. This design allows the \textrm{LF encoder} to be parameterized with higher complexity, as wind noise is more concentrated in the lower sub-bands, while the \textrm{HF encoder} uses relatively fewer parameters. Both encoders employ a flatten layer to reshape frequency and channel axes into one-dimensional vectors.
The output features of the low- and high-frequency encoder modules \textrm{LF encoder} and \textrm{HF encoder} are denoted as \(\mathbf{C}_l\) and \(\mathbf{C}_h\) respectively. Moreover, \(\mathbf{C}_l\) 
is further processed using a temporal GRU layer. Subsequently, the outputs are concatenated and passed to a fully connected layer with Sigmoid activation to estimate the intermediate real magnitude mask $\widebar{\mathbf{M}}_{m}$,
{\small
\begin{equation}
\begin{aligned}
[\mathbf{H}_l, \mathbf{C}_h] &= [\operatorname{GRU}_l(\mathbf{C}_l),\mathbf{C}_h] \\
\widebar{\mathbf{M}}_{m} &= \mathbf{W} [\mathbf{H}_l, \mathbf{C}_h] + \mathbf{b},
\end{aligned}
\end{equation}
}where \(\mathbf{W}\) and \(\mathbf{b}\) represent the learnable weight matrix and bias vector of the fully-connected (linear) layer, respectively. The mask $\widebar{\mathbf{M}}_{m}$ is used to compute intermediate complex features as described in \cite{ULCNet_24},
{\small
\begin{equation}
\widetilde{\mathbf{Y}}_{r} = \widebar{\mathbf{M}}_{m} \cdot \cos\left({\widetilde{\mathbf{X}}_{p}}\right) ~\mbox{and}~
\widetilde{\mathbf{Y}}_{i} = \widebar{\mathbf{M}}_{m} \cdot \sin\left({\widetilde{\mathbf{X}}_{p}}\right).
\label{eq:IntermediateFeature}
\end{equation}
}These intermediate features $\widetilde{\mathbf{Y}}_{r}$ and $\widetilde{\mathbf{Y}}_{i}$ are then concatenated along the channel dimension and passed through a lightweight CNN architecture to estimate a complex mask $\mathbf{M}$. The mask is used to reconstruct the power-law compressed estimate of either the desired clean signal $\widetilde{\mathbf{D}}$ (for \textit{wind rejection}) or the wind component $\widetilde{\mathbf{W}}$ (for \textit{wind extraction}) as follows,
{\small
\begin{equation}
\widetilde{\mathbf{D}} \text{ or } \widetilde{\mathbf{W}} = \widetilde{\mathbf{X}}_{m} \cdot \mathbf{M}_m \cdot e^{\jmath (\widetilde{\mathbf{X}}_{p} + \mathbf{M}_{p})},
\label{eq:CleanSpeechEst}
\end{equation}
}where $\mathbf{M}_m$ and $\mathbf{M}_p$ denote the magnitude and phase components of the complex-valued mask $\mathbf{M}$, and $\widetilde{\mathbf{X}}_{m}$ and $\widetilde{\mathbf{X}}_{p}$ are as defined in \eqref{eq:mag_phase}. Finally, after applying power-law decompression using Eq.~\ref{eq:powerlaw_real_imag} with a decompression factor $\beta = \frac{1}{\alpha}$, followed by inverse \ac{STFT}, we obtain the estimated time-domain desired signal $\mathbf{\hat{d}}[n]$ or wind noise $\mathbf{\hat{w}}[n]$.
\section{Experimental Setup}
\label{sec:results}

\subsection{Model Configuration and Baseline}
\textbf{\(\textrm{\emph{WindNetLite}}\)}: The architecture follows ULCNet\cite{ULCNet_24}-based convolutional recurrent network design with parallel encoder blocks. The channel-wise feature reorientation operates with \( L = 40 \) and an overlap factor of \( r = 0.4 \), resulting in \( N = 10 \) sub-bands. The input features to both the low-frequency (LF) and high-frequency (HF) encoders have a shape of \( (B, T, 40, 5) \). The LF encoder consists of four convolutional layers with \(\{32, 64, 96, 128\}\) filters. Since the higher frequency range does not require full resolution, an average pooling layer with a pool size and stride of \( (1, 2) \) is applied to the HF input before encoding. The HF encoder comprises three convolutional layers with \(\{8, 16, 64\}\) filters. All convolutions are performed with a kernel size of \( (1, 3) \) and a stride of \( (1, 2) \), except for the first layer, which uses a stride of \( (1, 1) \).
The extracted features are subsequently processed using point-wise convolution layers, with $32$ filters for LF and $16$ filters for HF, to produce the final convolutional features: \( \mathbf{C}_l \in \mathbb{R}^{B \times T \times 5 \times 32} \) and \( \mathbf{C}_h \in \mathbb{R}^{B \times T \times 5 \times 16} \), respectively. The LF feature \( \mathbf{C}_l \) is further processed through a temporal GRU block with $128$ units, yielding \( \mathbf{H}_l \in \mathbb{R}^{B \times T \times 128} \). These features are then concatenated and passed through a fully connected layer with 257 output units. In the second stage, the CNN consists of $2$ convolutional layers with $32$ filters each, followed by a point-wise convolution layer, which outputs the final complex mask.

\noindent\textbf{\(\textrm{\emph{WindNet}}\)}: As the baseline, we adapted the ULCNet model, which we refer to as \emph{WindNet}. In this work, we remove the bidirectional frequency GRU layer and reduce the number of temporal GRU layers in each sub-band from $2$ to $1$. These modifications are intended to provide a baseline model, \emph{WindNet}, with complexity comparable to the proposed \emph{WindNetLite}.

\begin{table*}[thbp!]
\centering
\small
\caption{Performance on the test set for scenarios with speech and instrumental music used as targets.}
\setlength{\tabcolsep}{3pt}
\begin{tabular} {l c c c c c c c c c c}
    \toprule
    \centering
    \multirow{2}{*}{\textbf{Processing}}
    & \multirow{2}{*}{PLF}
     & \multirow{2}{*}{Params} & \multirow{2}{*}{RTF (\(\%\))} &  \multirow{2}{*}{MHz} & \multicolumn{3}{c}{Speech} & \multicolumn{3}{c}{Ins. Music}     \\
     \cmidrule(lr){6-8}     
     \cmidrule(l){9-11}
     &  &  &  & & PESQ & SI-SDR & Leakage ($\mathcal{L}$)& PEAQ & SI-SDR & Leakage ($\mathcal{L}$) \\
    \midrule
    Noisy & - & - & - & - & 1.80 & 3.03 & - & -3.01 & 1.56 & -\\
    ULCNet & 0.3 & 688K & 0.127 & 182 MHz & 2.23 & 12.75 & -1.76 & - & - & -\\
    \emph{WindNet} (Rej.) & 0.3 & 365K & 0.061 & 87 MHz & \textbf{2.26} & 12.71 & -2.72 & \textbf{-2.32} & 6.65 & \textbf{-2.90}\\
    \emph{WindNet} (Ext.) & 1.0 & 365K & 0.061 & 87 MHz & \textbf{2.26} & 13.90 & -2.32 & -2.59 & 7.58 & -2.75\\
    \emph{WindNetLite} (Rej.) & 0.3 & \textbf{249K} & \textbf{0.051} & \textbf{73 MHz} & 2.24 & 12.27 & \textbf{-2.81} & -2.35 & 6.30 & -2.80\\
    \emph{WindNetLite} (Ext.) & 1.0 & \textbf{249K} & \textbf{0.051} & \textbf{73 MHz} & 2.23 & \textbf{14.31} & -2.43 & -2.57 & \textbf{7.67} & -2.60\\
    \bottomrule
\end{tabular}
\label{tab:Objectiveresult}
\vspace{-.3em}
\end{table*}
\vspace{-.5em}
\subsection{Training and Test Datasets}
\label{sec:training_data}
We trained the proposed \emph{WindNet} and \emph{WindNetLite} in both the rejection and extraction modes. To compose the training and test datasets, we recorded real wind noise in an outdoor environment using $20$ different microphones, corresponding to a total duration of 10 hours. For both training and validation datasets, we mixed the recorded wind noise with samples from the AudioSet dataset \cite{gemmeke2017audio} at signal-to-noise ratio (SNR) levels ranging from $-20$ to $20$ dB. AudioSet includes a diverse range of audio categories such as speech, music, and background noises (e.g., roadside traffic, barking, and chirping). The signals from the 'wind' category in AudioSet are excluded from the dataset composition, 
to prevent data leakage. 
The resulting training dataset had a total duration of approximately \unit[200]{hours}.

We created two test sets, each representative of a distinct target signal type commonly encountered in audio recording applications: speech and instrumental music. For the speech test set, the desired signals were taken from the VCTK Voice Bank corpus \cite{VCTK}, while the instrumental music samples were drawn from the MUSDB18 corpus \cite{musdb18}. These clean signals were mixed with previously unseen recorded wind noise at SNR levels randomly sampled from the set $\{-20,\textrm{dB}, -10,\textrm{dB}, 0,\textrm{dB}, 10,\textrm{dB}, 20,\textrm{dB}\}$. Each test set consisted of $200$ samples with a $10$~s duration per sample.

\begin{figure}[t!]
    \centering
    \centering
    \includegraphics[width=0.45\textwidth, trim=0 0 0 25pt, clip]{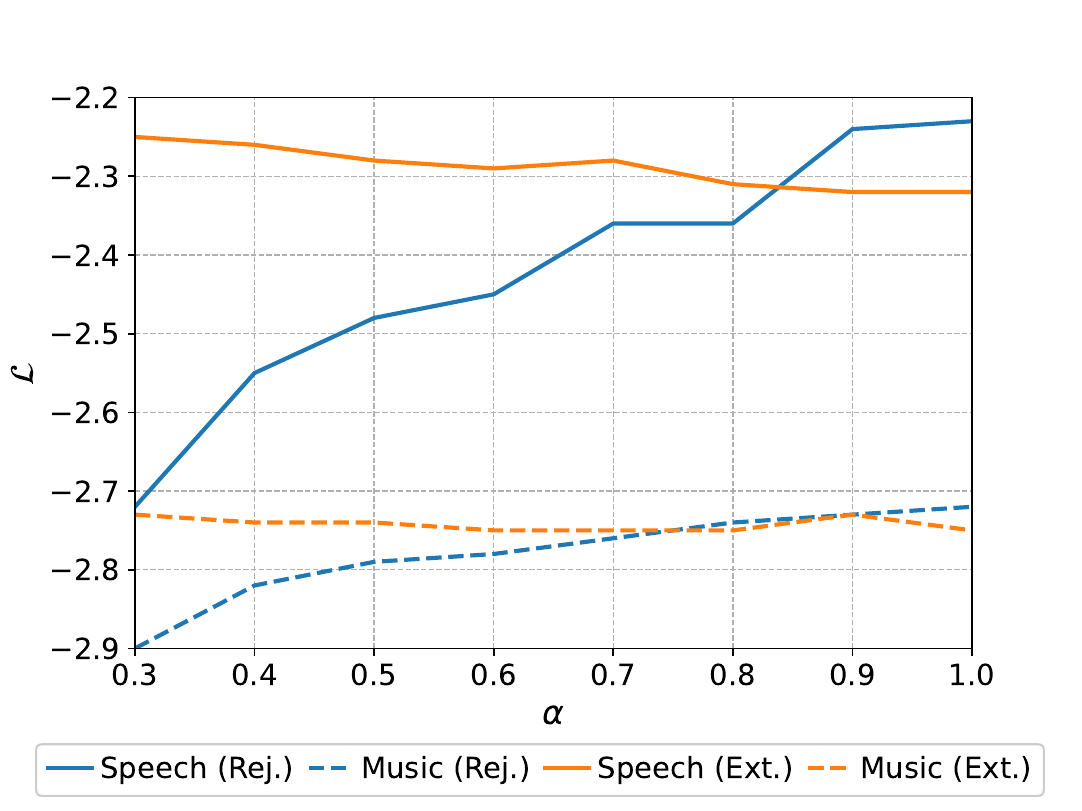} 
    \caption{Leakage $\mathcal{L}$ versus Power-law factor $\alpha$ for speech and instrumental music for extraction and rejection modes.}
    \label{fig:1}
\end{figure}
\subsection{Training Details}
We used an STFT with an FFT length of 512, and a Hann window length of $32$~ms, with $50\%$ overlap between successive frames, which resulted in $257$ frequency bins at $16$kHz sampling rate. The MSE computed between the ground truth signal STFT and its estimate computed by the DNN was used as the loss function, as is done in \cite{ULCNet_24} for noise reduction. The Adam optimizer was used with an initial learning rate (LR) of $0.0004$. The training framework also utilized a scheduler to reduce LR by a factor of $10$ every three epochs. The batch size was $400$ and each training example had a duration of $3$~s.

\subsection{Evaluation Metrics} To evaluate different methods, we used PEAQ \cite{thiede2000peaq}, PESQ \cite{rix2001perceptual} and SI-SDR \cite{le2019sdr} on the desired signal as the primary objective metrics. For a more detailed analysis of wind noise leakage, we employed the log-spectral distance (LSD) metric to compute the wind noise leakage metric \( \mathcal{L} \), defined as follows:
{\small
\begin{equation}
    \mathcal{L} \triangleq - \sqrt{
        \frac{1}{T F} \sum_{t=1}^{T} \sum_{f=1}^{F} 
        \left( \log \left| \widehat{D}_{t,f} \right| - \log \left| W_{t,f} \right| \right)^2,
    }
    \label{eq:lsd_leakage}
\end{equation}
}
where $\widehat{D}_{t,f}$ is the estimated desired signal complex STFT coefficient at time frame $t$ and frequency bin $f$, $W_{t,f}$ is the noise (wind) component coefficient, and $T$ and $F$ denote the total number of time frames and frequency bins, respectively. The negative sign before LSD is included to emphasize that a larger distance between the desired signal and noise corresponds to less leakage, indicating more effective wind noise reduction.

\section{Results}
\subsection{Impact of Power-Law Factor}\label{sec:powerlawRes}
Our preliminary experiments have shown that the power-law compression described in Section~\ref{sec:proposed_solution} significantly affects performance. Although the power law factor \(\alpha = 0.3\) was recommended in \cite{ULCNet_24} for speech enhancement, this value may not be optimal for \ac{WNR} depending on the target type, i.e., rejection or extraction.  Thus, we evaluated WNR for \(\alpha\) from $0.3$ to $1.0$ in $0.1$ increments for both modes. 
Figure~\ref{fig:1} shows the wind noise leakage \(\mathcal{L}\) as a function of \(\alpha\). For rejection, \(\alpha = 0.3\) achieves the lowest leakage, whereas \(\alpha = 1.0\) performs better for extraction. Informal listening tests confirmed these results, so we used \(\alpha = 0.3\) for rejection and \(\alpha = 1.0\) for extraction in all subsequent experiments.
\subsection{Performance Comparison}
\label{ssec:Objective results}
Table~\ref{tab:Objectiveresult} presents the \ac{WNR} performance of our proposed \emph{WindNetLite} and the baseline in both rejection and extraction modes, compared to the noisy signal.
On the speech test set, \emph{WindNet} achieves the highest PESQ score of $2.26$, while \emph{WindNetLite} matches ULCNet with a comparable score of $2.23$. In terms of SI-SDR, extraction method outperform the rejection method, both \emph{WindNet} and \emph{WindNetLite} yield higher SI-SDR values ($13.90$ and $14.31$), whereas in rejection mode they achieve lower leakage ($-2.72$ and $-2.81$). The ULCNet performs slightly lower than other models. However, it was trained for general speech enhancement rather than targeted \ac{WNR}.
On the instrumental music test set, rejection mode yields higher PEAQ scores ($-2.32$ for \emph{WindNet}, $-2.35$ for \emph{WindNetLite}). In extraction mode, \emph{WindNetLite} slightly outperforms \emph{WindNet} in SI-SDR, with scores of $7.67$~dB and $7.58$~dB, respectively.  Leakage metrics show that rejection methods outperform extraction, with scores of $-2.90$ for \emph{WindNet} and $-2.80$ for \emph{WindNetLite}. 
In informal listening, we observe that the extraction mode introduces less distortion to the target signal, while the rejection mode provides stronger wind suppression.  As a result,, the extraction method always achieves better SI-SDR results whereas the rejection method yields better leakage scores, since the leakage metric is more biased toward noise suppression.

In terms of computational complexity, both \emph{WindNet} and \emph{WindNetLite} significantly reduce computational costs compared to ULCNet, as evident from their parameter count and real-time factors (RTFs) measured on a single-core Cortex A53 1.43~GHz. As reported in Table~\ref{tab:Objectiveresult}, \emph{WindNet} achieves an RTF of $0.061$ ($87$ MHz), and \emph{WindNetLite} reaches $0.051$ ($73$ MHz) .
\begin{figure}[t]
    \centering
    \includegraphics[width=0.48\textwidth, trim={0 0pt 0 5pt},clip]{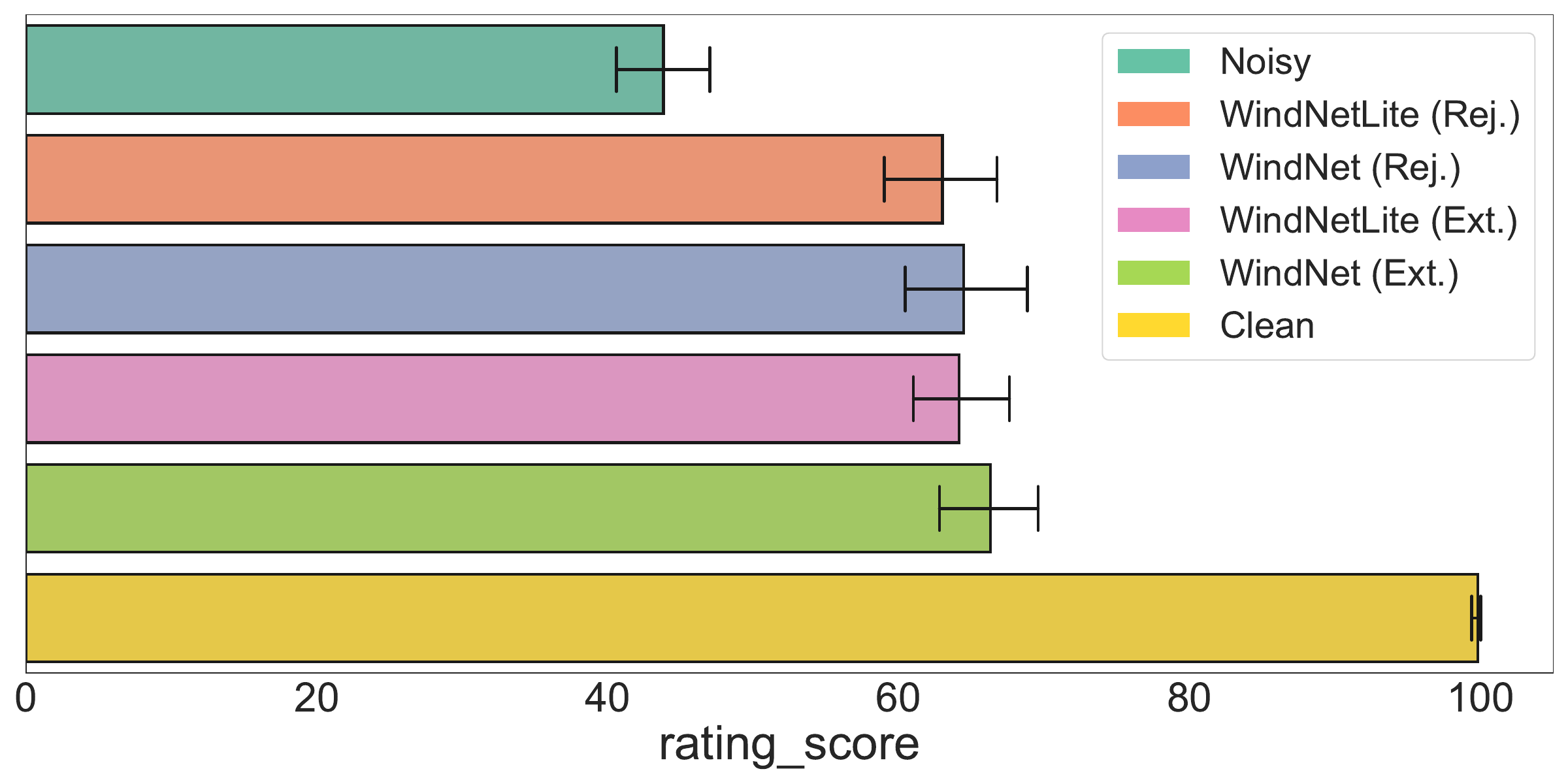} 
    \caption{Results of MUSHRA listening test}
    \label{fig:mushra}
    \vspace{-1.5em}
\end{figure}

\subsection{Subjective Evaluation}
\label{ssec:Listening tests}
To assess the subjective performance differences, we conducted a MUSHRA listening test. The test is carried out using the WebMUSHRA\cite{schoeffler2015towards} framework, and the listeners were asked to rate the overall performance based on the signal quality and introduced distortions.
In total, 10 listeners took part in the listening test, which consisted of $10$ samples with signal-to-noise ratios ranging from $-8$ to $-20$~dB. The listeners were asked to rate the clean signal as 100, which was provided as the reference. As illustrated in Fig. ~\ref{fig:mushra}, \emph{WindNet} in extraction mode attained the highest mean rating of $66.29$ while the other three models achieved comparable results. The listening samples can be found at  https://fhgwnr.github.io/AIWNR/. 

\section{Conclusions}
\label{sec:Conclusions}
We have shown that leveraging wind noise spectral characteristics enables designing a very low-complexity DNN model suitable for resource-constrained embedded devices. Our proposed \emph{WindNetLite} model requires only $40\%$ of the RTF of the SOTA ULCNet and $84\%$ of the baseline \emph{WindNet} model, while achieving comparable results. We also show that it is important to study different hyperparameters and signal reconstruction methods to achieve a better trade-off between signal distortion and suppression quality.


\balance
\small
\bibliographystyle{ieeetr}
\bibliography{references}


\end{document}